\documentclass[a4paper,fleqn,usenatbib]{mnras}

\usepackage{newtxtext,newtxmath}

\usepackage[T1]{fontenc}
\usepackage{ae,aecompl}

\usepackage{graphicx}	
\usepackage{amsmath}	
\usepackage{amssymb}	
\usepackage{dsfont}
\usepackage{relsize}

\newcommand\Nz{{N_{\mathsmaller{0}}}}
\newcommand\nn{\overline{N}}
\newcommand\ff{\overline{\mathbfss{F}}}
\newcommand\zz{\overline{\mathbfss{Z}}}
\newcommand\Sigmaz{{\boldsymbol{\Sigma}}_{\mathsmaller{0}}}

\newcommand\PPhi{\boldsymbol{\Phi}}

\newcommand\R{\mathbfss{R}}
\newcommand\G{\mathbfss{G}}
\newcommand\F{\mathbfss{F}}
\newcommand\Z{\mathbfss{Z}}

\newcommand\W{\mathbfss{W}}
\newcommand\Ss{\mathbfss{S}}
\newcommand\Ii{\mathbfss{I}}
\newcommand\SSigma{\boldsymbol{\Sigma}}
\newcommand\RGt{{\R_{\mathrm{grid}}}}
\newcommand\RFPhit{{\R_{\mathrm{sing}}}}

\DeclareMathOperator{\Diag}{Diag}

\title[Robust dimensionality reduction for interferometric imaging]{Robust dimensionality reduction for interferometric imaging of Cygnus A}
\author[S. Vijay Kartik et al.]{
S. Vijay Kartik,$^{1}$\thanks{E-mail: vijay.kartik@epfl.ch}
Arwa Dabbech,$^{2}$
Jean-Philippe Thiran,$^{1}$
and Yves Wiaux$^{2}$
\\
$^{1}$ Signal Processing Laboratory (LTS5), Ecole polytechnique f{\'e}d{\'e}rale de Lausanne (EPFL),
      CH-1015 Lausanne, Switzerland\\
$^{2}$Institute of Sensors, Signals, and Systems, Heriot-Watt University, Edinburgh EH14 4AS, UK
}

\date{Accepted XXX. Received YYY; in original form ZZZ}

\pubyear{2017}

\begin{document}
\label{firstpage}
\pagerange{\pageref{firstpage}--\pageref{lastpage}}
\maketitle

\begin{abstract}
Extremely high data rates expected in next-generation radio interferometers necessitate a fast and robust way to process measurements in a big data context.
Dimensionality reduction can alleviate computational load needed to process these data, in terms of both computing speed and memory usage.
In this article, we present image reconstruction results from highly reduced radio-interferometric data, following our previously proposed data dimensionality reduction method, $\RFPhit$, based on studying the distribution of the singular values of the measurement operator.
This method comprises a simple weighted, subsampled discrete Fourier transform of the dirty image.
Additionally, we show that an alternative gridding-based reduction method works well for target data sizes of the same order as the image size.
We reconstruct images from well-calibrated VLA data to showcase the robustness of our proposed method down to very low data sizes in a `real data' setting.
We show through comparisons with the conventional reduction method of time- and frequency-averaging, that our proposed method produces more accurate reconstructions while reducing data size much further, and is particularly robust when data sizes are aggressively reduced to low fractions of the image size.
$\RFPhit$ can function in a block-wise fashion, and could be used in the future to process incoming data by blocks in real-time, thus opening up the possibility of performing `on-line' imaging as the data are being acquired.
\textsc{matlab} code for the proposed dimensionality reduction method is available on GitHub.
\end{abstract}




\section{Introduction}
\label{sec:introduction}
The planned data processing pipelines for next generation radio interferometers like the Square Kilometre Array (SKA) are expected to witness an explosion of radio-interferometric data.
The estimated and oft-quoted data rate for the SKA is about five terabits per second~\citep{broekema_square_2015}. Such a data rate is necessary to enable the SKA to meet its ambitious science goals through very high-fidelity images of the radio sky, with gigapixel resolution and high dynamic ranges of six/seven orders of magnitude. Yet, it is very demanding to be handled on-the-fly with current imaging techniques.
Two major directions of research towards addressing the challenge of processing such high data rates are, namely, fast/efficient imaging and data size reduction.
Novel imaging methods based on sparse reconstruction are being actively developed to enable real-time processing to scale with the deluge of acquired data~\citep{wiaux_compressed_2009,li_application_2011,carrillo_purify:_2014,dabbech_moresane:_2015,ferrari_multi-frequency_2015,garsden_lofar_2015,onose_scalable_2016}.
These proposed imaging methods employ iterative, convex optimization-based algorithms, and are designed to exploit image sparsity in suitable bases.
These iterative reconstruction algorithms are often computationally intensive, hence requiring big data processing solutions in a distributed, high performance computing (HPC) setting.

In addition to HPC-ready imaging algorithms, another interesting direction is the development of dimensionality reduction methods, where large amounts of data are embedded to more manageable sizes on which further processing such as imaging and calibration is performed.
Naturally, the aim is to be able to process the embedded data and produce images with no loss in the reconstruction quality when compared with images produced from the full data set.
Dimensionality reduction pertaining to radio-interferometric data has traditionally been performed using time- and frequency- averaging, with averaging bins chosen according to a combination of factors including the desired data reduction, field of view, and relevant science objectives among others.
These averaging methods, however, can only reduce the data size to a limited extent, depending on the number of time and frequency data-points available respectively as snapshots and channels.
Moreover, this type of averaging introduces `smearing' artefacts in the reconstructed images.
Smearing presents in the image as attenuated off-centre sources.
This attenuation in itself may not always be undesirable, however.
For example, a decidedly fortunate effect of smearing is the attenuation of the global background known as far sidelobe confusion noise (FSCN).
In general, though, smearing artefacts in an image are detrimental to image quality, since the overall apparent flux is reduced and the point spread function is distorted~\citep{atemkeng_using_2016}.
The ill-effects of averaging visibilities are known and documented in the literature, and several approaches to mitigate them through windowing/filtering methods have been proposed, e.g. by~\citet{offringa_post-correlation_2012,parsons_optimized_2016}.
\citet{atemkeng_using_2016} proposed a baseline-dependent windowing method to minimize smearing artefacts, while continuing to use an averaging-based method for data size reduction.

Linear dimensionality reduction using random projections has also been studied in numerical algebra research, notably under the name of `sketching', which describes the process of solving a high-dimensional optimization problem by mapping it to a lower-dimensional subspace where solving the problem would be more reasonable~\citep{mahoney_randomized_2011,woodruff_sketching_2014}.
The feasibility of random projections to reduce data dimensionality was investigated by~\citet{vijay_kartik_dimension_2015}.
Further advances in this direction leading to a proposed Fourier dimensionality reduction model for interferometry are described in recent work by ~\citet{vijay_kartik_fourier_2017}.
The proposed method is presented as a fast and modular embedding operator consisting of taking a weighted, subsampled Fourier transform of the dirty image.

This article demonstrates the robustness of this recently proposed method down to very low data sizes in a real data setting, using well-calibrated observations of the radio galaxy Cygnus A.
Section~\ref{sec:background} provides the background of the ill-posed inverse problem of radio-interferometric imaging and briefly describes the sparse reconstruction-based approach to solving this problem.
In section~\ref{sec:dimred}, we present our proposed dimensionality reduction method based on the singular value decomposition (SVD) of the measurement operator and compare it with the existing standard dimensionality reduction method of averaging continuous visibilities over time and frequency.
We additionally present reconstruction results using a modified gridding-based reduction method, and make comparisons with our SVD-based method in the case of low data size.
The mathematical background and motivation for the proposed method can be found in~\citet{vijay_kartik_fourier_2017}.
Image reconstruction results using real radio-interferometric data are presented in section~\ref{sec:results}, showing the robustness of our reduction method and its effectiveness in a big data setting, owing to the achieved drastic reduction in data size.
We show that images reconstructed using our aggressive data dimensionality reduction method in combination with the convex optimization-based imaging algorithm proposed in~\citet{onose_scalable_2016} compare favourably to images obtained either with gridded visibilities or `classical' dimensionality reduction like averaging, and perform equally well or better when compared to images reconstructed from the full data set of continuous visibilities.
We conclude in section~\ref{sec:conclusions} with comments on the suitability of dimensionality reduction as a means of addressing the technical challenges of dealing with very high data rates, and indicate avenues for future work in real-time processing of SKA data.
\textsc{matlab} code for the reconstructions and different dimensionality reduction methods is available on GitHub\footnote{\url{http://basp-group.github.io/fourierdimredn}}.

\section{Radio-interferometric imaging}
\label{sec:background}
\subsection{Measurement equation}
Radio interferometers measure Fourier components of sky brightness.
This \emph{simplified} interpretation can be traced to the van Cittert-Zernike theorem~\citep{thompson_interferometry_2001}, which under the assumptions of (i) small field of view, and (ii) narrow-band incoherent signals, can be rephrased as the following ``measurement equation'' -
\begin{equation}
\label{eq:contrimeintegral}
\boldsymbol{y}(u,v) = \iint \limits_{\Omega} \boldsymbol{A}(l,m,u,v)\,\boldsymbol{x}(l,m)\operatorname{e}^{-2\pi\!\operatorname{i} (ul+vm)}\,\operatorname{d}\!l\operatorname{d}\!m,
\end{equation}
where $\boldsymbol{y}(u,v)$ are the `visibilities' obtained from an underlying sky brightness `image' $\boldsymbol{x}(l,m)$.
$\boldsymbol{A}(l,m,u,v)$ includes all antenna properties like collecting area, beam pattern and other direction dependent effects (e.g., the $w$-term).
The integral is computed over $\Omega = \{(l,m):l^2+m^2<1\}$.
($u$,$v$) form the coordinates of the visibilities $\boldsymbol{y}$ in the plane perpendicular to the line of sight, and ($l$, $m$) those of the sky brightness distribution $\boldsymbol{x}$.
A detailed discussion of the measurement equation can be found in~\citet{thompson_interferometry_2001,smirnov_revisiting_2011}.
The discretized form of the linear measurement model can then be written as
\begin{equation}
\label{eq:rime}
\boldsymbol{y}=\PPhi \boldsymbol{x} + \boldsymbol{n},
\end{equation}
where $\boldsymbol{x} \in \mathbb{R}^{N}$ is now the (vectorized) image and $\boldsymbol{y} \in \mathbb{C}^{M}$ the visibilities vector, corrupted with an additive measurement noise $\boldsymbol{n} \in \mathbb{C}^{M}$.
$\PPhi \in \mathbb{C}^{M\times N}$ is the measurement operator denoting the linear relation between the signal and the continuous visibilities, given by $\PPhi=\G\ff\boldsymbol{\mathsf{D}}_{\mathsmaller{\mathrm{R}}}\Z\boldsymbol{\mathsf{B}}$, where $\boldsymbol{\mathsf{B}}$ denotes the primary beam, $\Z$ the zero-padding operator on the image needed to compute the discrete Fourier transform of $\boldsymbol{x}$ on a finer sampling grid in the Fourier domain, and $\ff$ the discrete Fourier transform operator for the oversampled grid (we later use $\F$ to denote the non-oversampled, image-sized Fourier transform).
$\G$ is a convolution interpolation operator to map from the discrete frequency grid to the continuous $uv$ plane, and $\boldsymbol{\mathsf{D}}_{\mathsmaller{\mathrm{R}}}$ is a diagonal matrix that contains the reciprocal of the inverse Fourier transform of the interpolation kernel used in $\G$, to undo the effects of the convolution.
We assume $\boldsymbol{\mathsf{B}}=\Ii$, where $\Ii$ is identity.
We also assume that direction-dependent effects can be accounted for in this approach by allowing general interpolation kernels in each row of the matrix $\G$.
For convenience in notation, we also define the combined operator $\zz=\boldsymbol{\mathsf{D}}_{\mathsmaller{\mathrm{R}}}\Z$ and rewrite the measurement operator as
\begin{equation}
\label{eq:phidefn}
\PPhi=\G\ff\zz\quad \in \mathbb{C}^{M\times N}.
\end{equation}
In this article we assume complete knowledge of $\PPhi$ and therefore pre-calibrated continuous visibilities $\boldsymbol{y}$.
We note that $\G^{\dagger}$ is a `gridding' operator that maps continuous visibilities to the oversampled discrete Fourier grid.
The superscript `$^{\dagger}$' is used to denote the adjoint operator/matrix.

\subsection{Convex optimization-based imaging methods}
\label{subsec:convexopt}
The incompleteness of the Fourier coverage modelled in equation~(\ref{eq:contrimeintegral}) leads to an ill-posed inverse problem when solving for $\boldsymbol{x}$ in equation~(\ref{eq:rime}).
The `classical' method to recover the image $\boldsymbol{x}$ is the \textsc{clean} algorithm~\citep{hogbom_aperture_1974} or one of its variants like Cotton-Schwab-\textsc{clean}~\citep{schwab_relaxing_1984} or multiscale-\textsc{clean}~\citep{cornwell_multiscale_2008}.
Several recently proposed imaging algorithms for radio interferometry rely on convex optimization methods, and have been shown to be more suitable for reconstructing images from radio interferometry data~\citep{carrillo_purify:_2014,dabbech_moresane:_2015,garsden_lofar_2015,onose_scalable_2016}.
The common thread among them is to exploit the sparsity of either the image or its representation in a suitable basis.
A combined sparse representation-based imaging method was proposed by~\citet{carrillo_purify:_2014} as the \textsc{sara} algorithm, which uses the concept of \emph{average sparsity} over multiple bases simultaneously.
\citet{vijay_kartik_fourier_2017} use the \textsc{sara} algorithm and employ an Alternating Direction Method of Multipliers (\textsc{admm})-based minimizer to reconstruct images from simulated data, with both a full set of continuous visibilities and different sets of reduced visibilities.
In the current work, we reconstruct images from real data using the Primal-Dual Forward-Backward (\textsc{pdfb})-based minimizer with the \textsc{sara} algorithm.
This \textsc{pdfb} algorithm proposed by~\citet{onose_scalable_2016} was shown to be faster and more easily parallelizable when compared to other iterative minimization methods, and is therefore an appropriate choice for recovering images from high-dimensional data with the possibility of deploying an HPC-ready version in the future.

\section{Dimensionality reduction}
\label{sec:dimred}
Dimensionality reduction comprises embedding a high-dimensional vector $\boldsymbol{y} \in \mathbb{C}^{M}$ to a lower-dimensional vector $\boldsymbol{y}' \in \mathbb{C}^{M'}$ such that $M' \ll M$.
A general linear dimensionality reduction method would achieve this through an operator $\R \in \mathbb{C}^{M'\times M}$.
Applying such an operator $\R$ to the measurement equation defined in equation~(\ref{eq:rime}), we obtain the `embedded' inverse problem
\begin{equation}
\boldsymbol{y}'=\R\PPhi \boldsymbol{x}+\R\boldsymbol{n}.
\label{eq:embeddedrime}
\end{equation}
We note here that the choice of $\R$ directly affects the computation and application of the measurement operator, and therefore needs to be designed such that computing embedded visibilities multiple times in an iterative algorithm with the \emph{combined} measurement operator $\PPhi'=\R\PPhi$ remains computationally efficient.

\subsection{Singular value-based dimensionality reduction}
\label{subsec:rsing}
\citet{vijay_kartik_fourier_2017} show that a fast and easily realizable dimensionality reduction of continuous visibilities can be achieved by taking into account the distribution of the singular values of the measurement operator $\PPhi$, and propose a Fourier dimensionality reduction model, given by an embedding operator
\begin{equation}
\label{eq:rfphit}
\R_{\mathrm{sing}}=\Sigmaz^{-1}\Ss\F\PPhi^{\dagger} \quad \in \mathbb{C}^{\Nz\times M},
\end{equation}
where $\PPhi$ is given by equation~(\ref{eq:phidefn}), $\SSigma$ is a diagonal matrix containing the singular values of $\PPhi$, and can be calculated to a good approximation by simply computing the square root of $\Diag(\F\PPhi^{\dagger}\PPhi\F^{\dagger})$, and $\Ss$ is a subsampling matrix to select the dimensions corresponding to the $\Nz \leq N$ non-zero singular values present in $\SSigma$ (giving us $\Sigmaz$), thus producing a dimensionality reduction below image size.

The full measurement operator, given by
\begin{equation}
\PPhi'_{\mathrm{sing}}=\Sigmaz^{-1}\Ss\F\PPhi^{\dagger}\PPhi \quad \in \mathbb{C}^{\Nz\times N},
\label{eq:phising}
\end{equation}
is suitably fast for repeated application in iterative minimization algorithms.
The speed of applying the full measurement operator $\PPhi'_{\mathrm{sing}}$ results from the simplicity of this dimensionality reduction, which consists of computing the dirty image and taking its discrete Fourier transform, and finally subsampling and weighting the result to form a low-dimensional data vector.

$\RFPhit$, in the above form, preserves all the information from the original measurement operator through preserving its null space by retaining all dimensions with non-zero singular values.
We may, however, be able to afford a more aggressive dimensionality reduction by using the subsampling matrix $\Ss$ to not only discard the zero-valued singular values, but also those that have a `small' magnitude (where `small' is relative to a threshold decided in advance).
Indeed, preliminary results on simulated data reported in~\citet{vijay_kartik_fourier_2017} show promising image reconstruction with very low data sizes.
We show with tests on real data that a trade-off between reaching very low data sizes and maintaining reasonable image quality can be achieved with $\RFPhit$, which is robust to discarding dimensions.

\subsubsection*{`On-line' dimensionality reduction}
\label{subsec:onlinedimred}
One issue with the current formulation of $\RFPhit$ stems from the forward model as described in equation~(\ref{eq:rime}).
It requires that the full data set $y \in \mathbb{C}^M$ be available to us initially, which implies handling $M$-dimensional data for the first set of computations, which is precisely what dimensionality reduction aims to avoid in the first place.
Furthermore, applying $\RFPhit$ would involve either computing the dirty image from high-dimensional data, or avoiding the higher dimension by pre-computing $\PPhi^{\dagger}\PPhi$ (or more precisely, $\G^{\dagger}\G$).
This pre-computation, even though it only needs to be performed once, requires handling $M$-dimensional structures yet again.
This seems to counteract all the computational savings that $\RFPhit$ claims to provide.
However, the block-separable structure of $\RFPhit$ can be exploited to resolve this issue.
The holographic matrix $\boldsymbol{\mathsf{H}}=\G^{\dagger}\G$~\citep{sullivan_fast_2012} combines the steps of computing and gridding visibilities to the discrete Fourier grid into a single, pre-computed mapping.
This can be split into separate blocks $\boldsymbol{\mathsf{H}}_{\mathrm{i}}$, giving
\begin{equation}
\boldsymbol{\mathsf{H}} = \G^{\dagger}\G = \sum{{\G_{\mathrm{i}}}^{\dagger}\G_{\mathrm{i}}} = \sum{\boldsymbol{\mathsf{H}}_{\mathrm{i}}} \quad \in \mathbb{C}^{o^2N\times o^2N}.
\label{eq:splitholographicmatrix}
\end{equation}
The discrete Fourier grid contains $o^2N$ points for an oversampling factor of $o$.
We note that the combined measurement operator $\PPhi'_{\mathrm{sing}}$ given in equation~(\ref{eq:phising}) can then be expressed as a sum of separate block-wise operators, giving
\begin{equation}
\PPhi'_{\mathrm{sing}} = \Sigmaz^{-1}\Ss\sum{\PPhi'_{\mathrm{i}}}\quad \in \mathbb{C}^{\Nz\times N},
\label{eq:phisingsum}
\end{equation}
where
\begin{equation}
\PPhi'_{\mathrm{i}} = \F{\PPhi_{\mathrm{i}}}^{\dagger}\PPhi_{\mathrm{i}} = \F\zz^{\dagger}\ff^{\dagger}\boldsymbol{\mathsf{H}}_{\mathrm{i}}\ff\zz\quad \in \mathbb{C}^{N\times N},
\end{equation}
$\PPhi_{\mathrm{i}}$ being blocks of the original measurement operator $\PPhi$.
We see here that the forward modelling is performed in the lower dimension $\Nz$ through block-wise measurement operators also in the lower dimension $N$, without ever performing computations in the higher dimension $M$ as outlined originally in equation~(\ref{eq:phidefn}).
We can thus apply the combined measurement operator in blocks $\PPhi'_{\mathrm{i}}$, leading to the reduced data vector $\boldsymbol{y}'_{\mathrm{i}}$, given by
\begin{equation}
\boldsymbol{y}' = \Sigmaz^{-1}\Ss\sum{\boldsymbol{y}'_{\mathrm{i}}}\quad \in \mathbb{C}^{\Nz},
\label{eq:yprimesum}
\end{equation}
where
\begin{equation}
\boldsymbol{y}'_{\mathrm{i}} = \F{\PPhi_{\mathrm{i}}}^{\dagger}\boldsymbol{y}_{\mathrm{i}}\quad \in \mathbb{C}^{N}.
\label{eq:batchdimredn}
\end{equation}
So, as each batch of visibilities $\boldsymbol{y}_{\mathrm{i}}$ is acquired, we can partially apply our dimensionality reduction $\RFPhit$ \emph{on-line} by taking a Fourier transform of the dirty image obtained from the batch.
This results in an immediate size reduction down to $N$, the image size -- without intermediate steps in the higher dimension $M$.
These $N$-sized data can be added sequentially as each batch of data is processed.
As a final step after all visibilities are acquired and reduced, we can apply the subsampling and weighting through $\Sigmaz^{-1}\Ss$, to obtain the low-dimensional embedded data that can be fed into imaging algorithms.
The data size at this stage would then be lower than image size.

Calculating $\Sigmaz^{-1}$ is crucial as these weights define the importance of the singular values of the measurement operator $\PPhi$, and are key in maintaining the reconstruction quality from embedded visibilities.
As outlined in the above equations, $\Sigmaz^{-1}$ and $\Ss$ can be applied at the end of a two-step reduction process, after reducing batches of data to image size $N$ on-line.
However, given our prior knowledge of $\PPhi$ -- which covers the telescope characteristics and the observation details for a given coverage -- we can compute $\Sigmaz^{-1}$ and $\Ss$ in advance, and apply them as part of the block-wise size reduction, thus further reducing data dimensionality to $\Nz$ per batch, instead of $N$.
One of the main advantages of this on-line method of dimensionality reduction is that the full-sized measurement operator $\PPhi$ never needs to be created or handled, thus saving computational resources.

With this scheme of applying the dimensionality reduction $\RFPhit$ to batches of data, we propose an avenue to handle high-dimensional data \emph{as they are acquired}.
This can potentially be plugged in as a module in the data processing pipeline, leading to an imaging step with already reduced data, while guaranteeing that the information content from the original data is retained.

\subsection{Gridding-based dimensionality reduction}
\label{subsec:rgrid}
Another dimensionality reduction we discuss in this article is derived from the standard method of `gridding' visibilities, i.\,e., mapping continuous visibilities to the discrete Fourier grid.
This can be modelled as applying $\R=\G^{\dagger}$.
Details of the mathematical background for this modelling are described in~\citet{vijay_kartik_fourier_2017}, where $\G^{\dagger}$ is also shown to provide a reasonable compromise between data size reduction and image reconstruction quality while using simulated data.
We maintain comparability with $\RFPhit$ by defining a modified version of the gridding operation as follows:
\begin{equation}
\label{eq:rgt}
\RGt=\overline{\W}_{\mathsmaller{0}}^{-1}\overline{\Ss}\G^{\dagger} \quad\in \mathbb{C}^{\nn\times M}.
\end{equation}
So, instead of the weighted, subsampled Fourier transform of the dirty image as in the case of $\RFPhit$ equation~(\ref{eq:rfphit}), here we implement a weighted, subsampled gridding of the continuous visibilities.
The weights $\overline{\W}$ are obtained by computing the square root of $\Diag(\G^{\dagger}\G)$, and the subsampling matrix $\overline{\Ss}$ selects the $\nn \leq o^2N$ dimensions corresponding to non-zero values of $\overline{\W}$ (resulting in $\overline{\W}_{\mathsmaller{0}}$), i.e., the $\nn$ discrete Fourier grid points that do have contributions from interpolation kernels.

Similarly to the aggressive dimensionality reduction with $\RFPhit$, we can also perform a further dimensionality reduction with $\RGt$.
From an initial mapping to the size of the oversampled Fourier grid, a conservative reduction discards dimensions corresponding to `holes' in the Fourier grid, denoting discrete grid points that are not covered by an interpolation kernel in the $uv$ plane.
For a further reduction in dimensionality, discrete Fourier grid points with minimal contributions from any interpolation kernels can be discarded.
As we discuss later in section~\ref{sec:results}, this is closely linked to the initial $uv$ coverage, and further dimensionality reduction may or may not be possible without losing reconstruction quality, depending on the location of the $uv$ points with respect to the discrete Fourier grid.

$\RGt$ can also be applied on-the-fly to data as they are acquired in batches.
We can follow the argument outlined in equations~(\ref{eq:phisingsum})-(\ref{eq:batchdimredn}), and use the block sub-structure of the holographic matrix $\boldsymbol{\mathsf{H}}$ (as defined in equation~(\ref{eq:splitholographicmatrix})) in a similar fashion, giving
\begin{equation}
\PPhi'_{\mathrm{grid}} = \overline{\W}_{\mathsmaller{0}}^{-1}\overline{\Ss}\sum{\PPhi'_{\mathrm{i}}} \quad\in \mathbb{C}^{\nn\times N},
\end{equation}
where
\begin{equation}
\PPhi'_{\mathrm{i}} = {\G_{\mathrm{i}}}^{\dagger}\PPhi_{\mathrm{i}} = \boldsymbol{\mathsf{H}}_{\mathrm{i}}\ff\zz \quad\in \mathbb{C}^{o^2N\times N}.
\label{eq:griddingblockphiprime}
\end{equation}
In this case, the reduced data vector $\boldsymbol{y}'_{\mathrm{i}}$ is then given by
\begin{equation}
\boldsymbol{y}' = \overline{\W}_{\mathsmaller{0}}^{-1}\overline{\Ss}\sum{\boldsymbol{y}'_{\mathrm{i}}} \quad\in \mathbb{C}^{\nn},
\end{equation}
where
\begin{equation}
\boldsymbol{y}'_{\mathrm{i}} = {\G_{\mathrm{i}}}^{\dagger}\boldsymbol{y}_{\mathrm{i}} \quad\in \mathbb{C}^{o^2N}.
\label{eq:griddingbatchdimredn}
\end{equation}
Data can be acquired in batches, and each batch $\boldsymbol{y}_{\mathrm{i}}$ can be immediately embedded to the oversampled discrete Fourier grid by applying $\G_{\mathrm{i}}^{\dagger}$.
This reduces the data dimensionality as a first step. 
Further reduction can possibly be applied by subsampling from this reduced data set.
We note from equations~(\ref{eq:griddingblockphiprime}) and~(\ref{eq:griddingbatchdimredn}) that the reduction is achieved without ever performing computations in the higher dimension $M$ as modelled in the original measurement operator given by equation~(\ref{eq:phidefn}).
Moreover, as for $\RFPhit$, $\overline{\W}_{\mathsmaller{0}}^{-1}$ and $\overline{\Ss}$ may be computed in advance and applied to each batch of data, thus reducing data size to $\nn$ instead of $o^2N$ from the beginning, instead of a two step process.

\subsection{Visibility averaging}
\label{subsec:averaging}
The conventional method of reducing data dimensionality in radio interferometry today is time- and frequency- averaging of continuous visibilities.
Time-averaging refers to averaging, across consecutive snapshots, the visibilities that correspond to the same baseline.  
Increasing the number of snapshots that one includes in an averaging bin leads to a bigger reduction in data size but comes at the cost of a coarser and more inaccurate coverage of the $uv$ space.
Frequency-averaging is performed across channels, averaging over visibilities corresponding to the same baseline for a given snapshot.
In the case of narrow-bandwidth channels in an averaging bin, the reduced data may remain a good approximation of the original data, and is indeed a quick and easy dimensionality reduction method.
Time- and frequency-averaging, however, have limitations.
Due to the limited number of snapshots in typical data sets, time-averaging cannot lead to drastic data dimensionality reduction. 
A major cause of loss of reconstruction quality, however, is the fact that time- and frequency-averaging are typically performed without being appropriately modelled in the measurement operator that is ultimately used for image reconstruction.
The measurement operator does not take into account the averaging operation performed, relying only on (now inaccurate) degridding kernels over the Fourier grid which do not correspond to the `reduced' data.
An appropriately modelled `new' measurement operator would be $\PPhi' = \R_{\mathrm{avg}}\PPhi$ where $\R_{\mathrm{avg}}$ would implement the averaging over continuous visibilities.
Instead, standard practice is to continue using the measurement operator $\PPhi$ in the imaging process.
The effects of this mis-modelling -- and indeed of the side-effects of the averaging itself -- can be seen in lower image reconstruction quality.
In particular, the effect of averaging over identical bins in all baselines is seen in reconstructed images in the form of `smearing'.
\citet{atemkeng_using_2016} propose baseline-dependent windowing functions to mitigate smearing effects, and further suggest that choosing larger time-averaging bins for shorter baselines (and vice versa) would reduce smearing in the image domain.

Time- and frequency- averaging cannot be performed indefinitely to reach arbitrarily low data sizes.
The absolute minimum reachable sizes are governed by initial conditions of the data acquisition, mainly the time intervals between snapshots, the number of channels and the overall $uv$ coverage.
In addition, averaging data to achieve very low data sizes may lead to a decline in the reconstruction quality -- both in itself and with respect to other data reduction methods.
Results of image recovery tests with extremely low-sized data support this conjecture, and visual comparisons between images recovered using visibilities by applying $\RFPhit$, $\RGt$ and simple averaging are shown in section~\ref{sec:results}.

Averaging lends itself readily to on-the-fly application.
Indeed, on-line batch processing of acquired data would be the simplest method of data reduction through averaging.
The ease of using batch-wise averaging is, however, tempered by the loss in image reconstruction quality that accompanies it.
This is particularly relevant for averaging aggressively to reach lower data sizes, as we show through image reconstruction results on real data.

\section{Image reconstruction results}
\label{sec:results}
\subsection{Data set details}
\label{subsec:vladata}

To test and compare the different data reduction methods described in section~\ref{sec:dimred}, we consider real data sets of observations of the radio galaxy Cygnus A.
The data consist of complex visibilities acquired as part of wideband observations in 2015-2016 by the Karl G. Jansky Very Large Array (VLA), operated by the National Radio Astronomy Observatory (NRAO) in New Mexico, USA.
The data correspond to the `C' band, over a narrow spectral window of 128\,MHz acquired over 64 2\,MHz wide channels, centred at 6680\,MHz.
Measurements were recorded using the VLA in configuration C, pointing at the phase centre given by RA$=19\rm{h}~59\rm{mn}~28.356\rm{s}$ ($J2000$) and DEC$=+40^{\circ}44\arcmin2.07\arcsec$.

\begin{figure}
	\centering
	\includegraphics[trim={0px 0px 0px 0px}, clip, width=0.95\columnwidth]{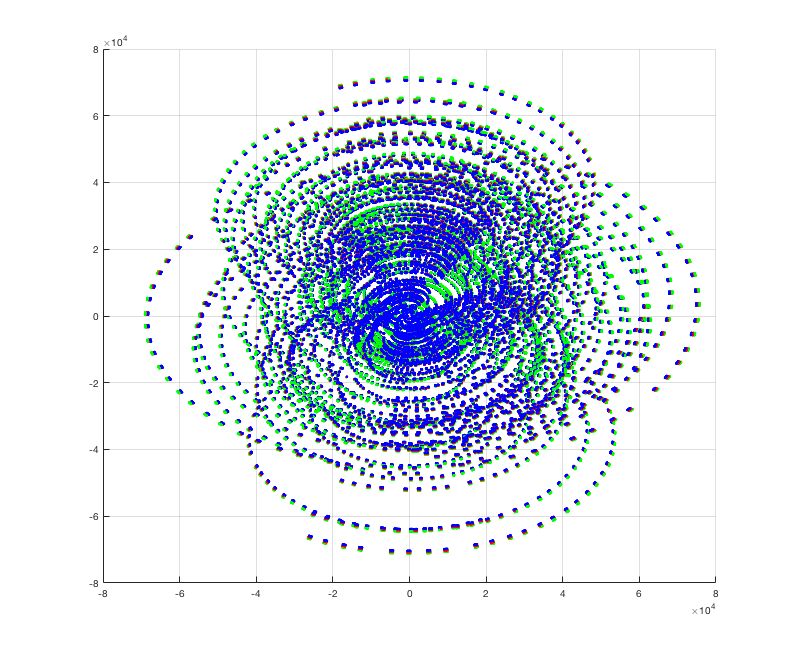}    
    \caption{Illustration of VLA $uv$ coverage used in the tests. Visibilities from different channels are collated to cover the same $uv$ plane. The three colours shown here signify three different channels. All $uv$ points were normalised to the maximum baseline. The values shown here are before scaling to lie between $[-\pi, \pi]$, to keep individual channel data visually distinct in spite of the narrow spectral window of the data set.}
    \label{fig:vlacoverage}
\end{figure}

Testing dimensionality reduction methods requires high-dimensional data.
For the considered data set, given the relatively small number of data points per channel ($\approx2\times10^5$) and the very narrow spectral window of observations, we decided to collate data from several channels together to form one single $uv$ coverage, from which the aim is to recover a single image.
Collating data from all 64 channels, however, was impractical due to computational limitations on (i) reconstructing without dimensionality reduction, and (ii) pre-computing the holographic matrix $\boldsymbol{\mathsf{H}}$ to enable application of $\RGt$ and $\RFPhit$.
We note here that this issue can be avoided in the future by applying on-line dimensionality reduction, which would ensure that we never handle the full data set, instead always taking per-block data as input, leading to manageable data sizes at each step of the imaging process.
Therefore, we chose 10 separate channels between 6630~MHz and 6720~MHz and concatenated their visibilities together, yielding a $uv$ coverage with about $2\times10^6$ data points.
Since the spectral slope in the data set was mild enough to be negligible over the observed bandwidth, we performed single frequency imaging and did not treat different channels separately as one would for hyperspectral imaging.
The $uv$ points were normalized to the maximum baseline and subsequently scaled to lie within $[-\pi,\pi]$.
An illustration of the $uv$ coverage is shown in Fig.~\ref{fig:vlacoverage}, with visibilities from three channels.

\subsection{Image recovery from VLA data}
\label{subsec:vlaresults}
The VLA data was used to reconstruct $256\times 256$ images with a pixel width of 0.5\arcsec -- this corresponds to recovering a signal of up to $2.5$ times the band-limit of the observations.
The full data set of $2\times10^6$ continuous visibilities is thus $\approx30$ times larger than the size of reconstructed images ($\approx6.5\times10^4$).
Images were reconstructed from data of varying dimensions, obtained through one of the following methods: (i) full data with no dimensionality reduction, (ii) simple averaging over time and frequency bins, (iii) dimensionality reduction by applying $\RGt$, and (iv) dimensionality reduction by applying $\RFPhit$.
In the first two cases, image recovery was performed using both \textsc{ms-clean} and \textsc{pdfb} algorithms.
In the latter two cases, only \textsc{pdfb} was used.

\begin{figure}
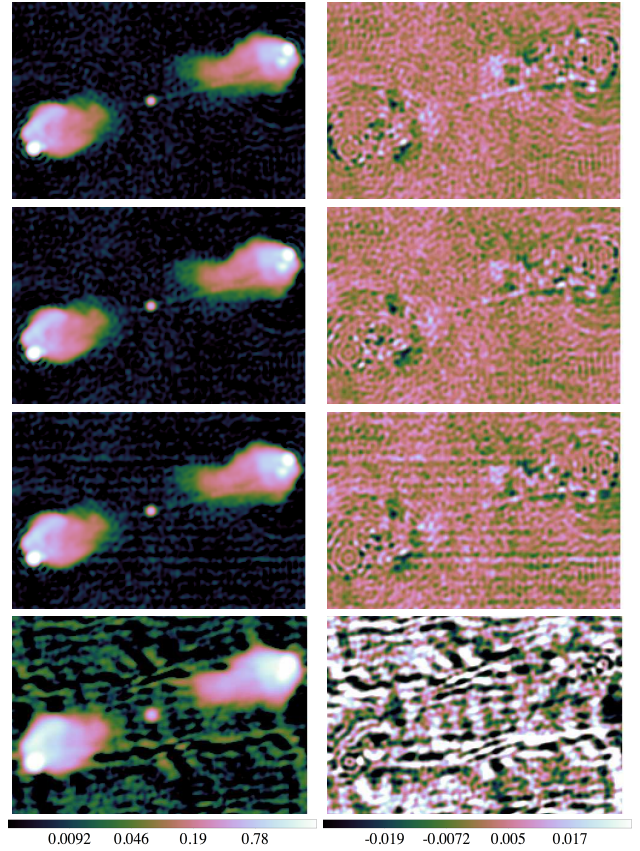

\centering
    \includegraphics[trim={150px 110px 100px 60px}, clip, width=0.48\columnwidth]{{{figs/i256.dl25.k100.hyperspec.q6-5-51.allvisibs.msclean.rec}}}
    	\vspace{0.03in}
    \includegraphics[trim={150px 110px 100px 60px}, clip, width=0.48\columnwidth]{{{figs/i256.dl25.k100.hyperspec.q6-5-51.allvisibs.msclean.resA}}}
    \vspace{0.03in}
    \includegraphics[trim={150px 110px 100px 60px}, clip, width=0.48\columnwidth]{{{figs/i256.dl25.k400.hyperspec.q6-5-51.averaging.msclean.rec}}}
    \includegraphics[trim={150px 110px 100px 60px}, clip, width=0.48\columnwidth]{{{figs/i256.dl25.k400.hyperspec.q6-5-51.averaging.msclean.resA}}}
    \vspace{0.03in}
    \includegraphics[trim={150px 110px 100px 60px}, clip, width=0.48\columnwidth]{{{figs/i256.dl25.k100.hyperspec.q6-5-51.averaging.msclean.rec}}}
    \includegraphics[trim={150px 110px 100px 60px}, clip, width=0.48\columnwidth]{{{figs/i256.dl25.k100.hyperspec.q6-5-51.averaging.msclean.resA}}}
    \vspace{0.03in}
    \includegraphics[trim={150px 110px 100px 60px}, clip, width=0.48\columnwidth]{{{figs/i256.dl25.k020.hyperspec.q6-5-51.averaging.msclean.rec}}}
    \includegraphics[trim={150px 110px 100px 60px}, clip, width=0.48\columnwidth]{{{figs/i256.dl25.k020.hyperspec.q6-5-51.averaging.msclean.resA}}}
    \vspace{0.1in}
	\includegraphics[trim={0px 0px 0px 0px}, clip, width=0.48\columnwidth]{{{figs/colourbar_rec}}}
    \includegraphics[trim={0px 0px 0px 0px}, clip, width=0.48\columnwidth] {{{figs/colourbar_resA}}}
    \caption[Reconstructed images using VLA data with \textsc{ms-clean}]{\textsc{ms-clean} image reconstructions using averaged visibilities. Left column: restored images in $\log_{10}$ scale. Right column: Briggs weighted residual images in linear scale. Rows denote final data size achieved after visibility averaging -- from top to bottom, $30N\approx 2\,000\,000$ visibilities ($\equiv$ full data, no averaging), $4N\approx260\,000$ visibilities (time-averaging over 10 snapshots), $N\approx65\,000$ visibilities (time-averaging over 20 snapshots), and $0.2N\approx13\,000$ visibilities (time-averaging over 30 snapshots and frequency-averaging over 10 channels). \textsc{ms-clean} was run with Briggs weighting (robust weighting parameter set to $-1$).}
    \label{fig:cleanresults}
\end{figure}

\begin{figure*}
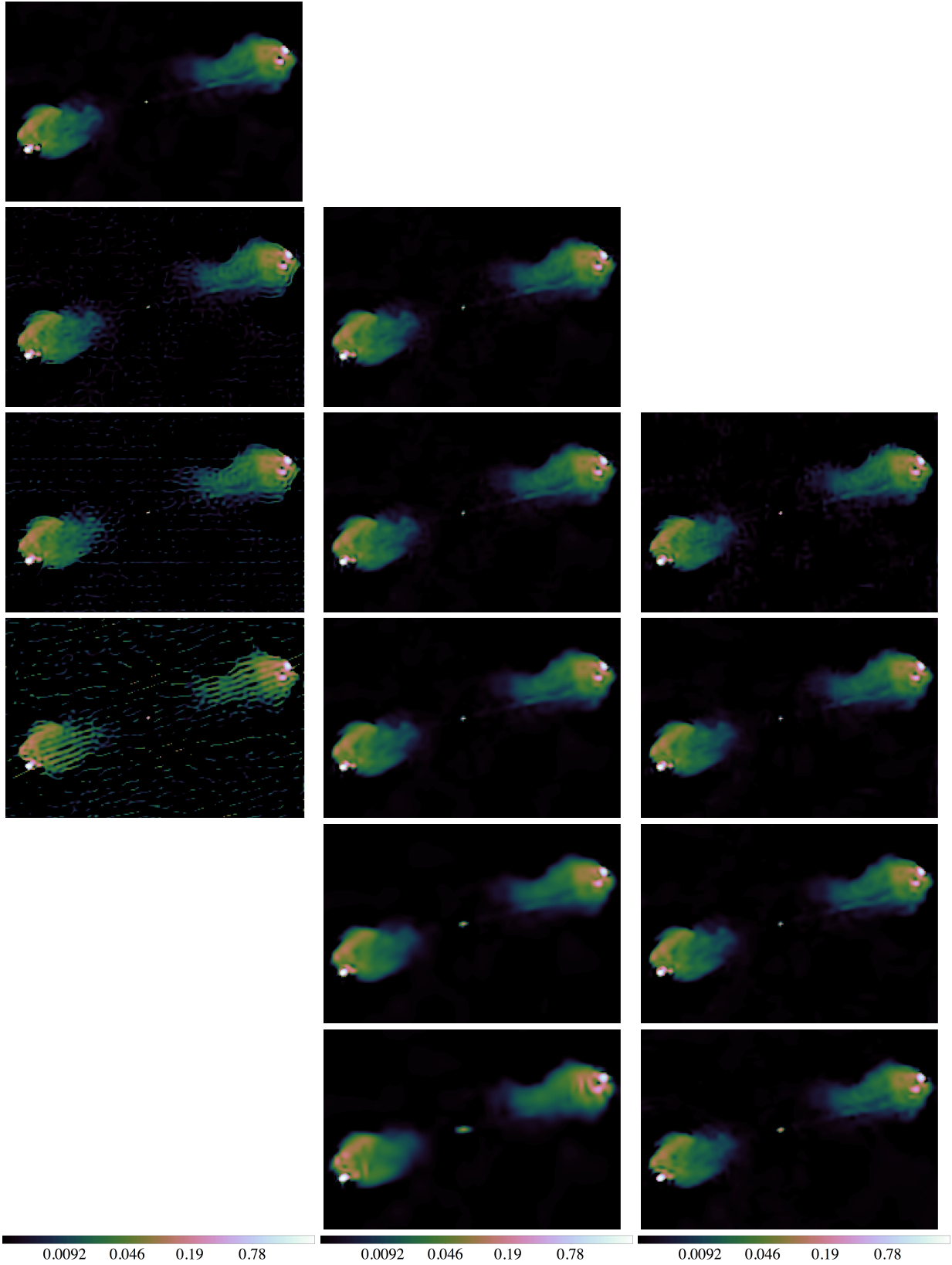

    \centering
    \includegraphics[trim={150px 110px 100px 60px}, clip, width=0.3\linewidth]{{{figs/i256.dl25.k100.n10000.eps6469.gamma3.hyperspec.q6-5-51.allvisibs.rec}}}
    \vspace{0.03in}
    \phantom{\includegraphics[trim={150px 110px 100px 60px}, clip, width=0.3\linewidth]{{{figs/i256.dl25.k100.n10000.eps6469.gamma3.hyperspec.q6-5-51.allvisibs.rec}}}}
    \phantom{\includegraphics[trim={150px 110px 100px 60px}, clip, width=0.3\linewidth]{{{figs/i256.dl25.k100.n10000.eps6469.gamma3.hyperspec.q6-5-51.allvisibs.rec}}}}
    \vspace{0.03in}
    \includegraphics[trim={150px 110px 100px 60px}, clip, width=0.3\linewidth]{{{figs/i256.dl25.k400.n05000.eps6170.gamma5.hyperspec.q6-5-51.averaging.rec}}}
    \includegraphics[trim={150px 110px 100px 60px}, clip, width=0.3\linewidth]{{{figs/i256.dl25.k400.n05000.eps6400.gamma5.hyperspec.q6-5-51.rgrid.rec}}}
	\phantom{\includegraphics[trim={150px 110px 100px 60px}, clip, width=0.3\linewidth]{{{figs/i256.dl25.k400.n05000.eps6400.gamma5.hyperspec.q6-5-51.rgrid.rec}}}}
    \vspace{0.03in}
    \includegraphics[trim={150px 110px 100px 60px}, clip, width=0.3\linewidth]{{{figs/i256.dl25.k100.n05000.eps2013.gamma5.hyperspec.q6-5-51.averaging.rec}}}
    \includegraphics[trim={150px 110px 100px 60px}, clip, width=0.3\linewidth]{{{figs/i256.dl25.k100.n05000.eps6400.gamma5.hyperspec.q6-5-51.rgrid.rec}}}
    \includegraphics[trim={150px 110px 100px 60px}, clip, width=0.3\linewidth]{{{figs/i256.dl25.k100.n05000.eps40.gamma5.hyperspec.q6-5-51.rsing.rec}}}
    \vspace{0.03in}
    \includegraphics[trim={150px 110px 100px 60px}, clip, width=0.3\linewidth]{{{figs/i256.dl25.k020.n05000.eps11038.gamma5.hyperspec.q6-5-51.averaging.rec}}}
    \includegraphics[trim={150px 110px 100px 60px}, clip, width=0.3\linewidth]{{{figs/i256.dl25.k020.n05000.eps6400.gamma5.hyperspec.q6-5-51.rgrid.rec}}}
    \includegraphics[trim={150px 110px 100px 60px}, clip, width=0.3\linewidth]{{{figs/i256.dl25.k020.n05000.eps20.gamma5.hyperspec.q6-5-51.rsing.rec}}}
    \vspace{0.03in}
	\phantom{\includegraphics[trim={150px 110px 100px 60px}, clip, width=0.3\linewidth]{{{figs/i256.dl25.k005.n05000.eps15.gamma5.hyperspec.q6-5-51.rsing.rec}}}}
	\includegraphics[trim={150px 110px 100px 60px}, clip, width=0.3\linewidth]{{{figs/i256.dl25.k005.n05000.eps6400.gamma5.hyperspec.q6-5-51.rgrid.rec}}}
	\includegraphics[trim={150px 110px 100px 60px}, clip, width=0.3\linewidth]{{{figs/i256.dl25.k005.n05000.eps15.gamma5.hyperspec.q6-5-51.rsing.rec}}}
    \vspace{0.03in}
	\phantom{\includegraphics[trim={150px 110px 100px 60px}, clip, width=0.3\linewidth]{{{figs/i256.dl25.k002.n05000.eps15.gamma5.hyperspec.q6-5-51.rsing.rec}}}}
	\includegraphics[trim={150px 110px 100px 60px}, clip, width=0.3\linewidth]{{{figs/i256.dl25.k002.n05000.eps6400.gamma5.hyperspec.q6-5-51.rgrid.rec}}}
    \includegraphics[trim={150px 110px 100px 60px}, clip, width=0.3\linewidth]{{{figs/i256.dl25.k002.n05000.eps15.gamma5.hyperspec.q6-5-51.rsing.rec}}}
    \vspace{0.03in}
	\includegraphics[trim={0px 0px 0px 0px}, clip, width=0.3\linewidth]{{{figs/colourbar_rec}}}
    \includegraphics[trim={0px 0px 0px 0px}, clip, width=0.3\linewidth]{{{figs/colourbar_rec}}}
    \includegraphics[trim={0px 0px 0px 0px}, clip, width=0.3\linewidth]{{{figs/colourbar_rec}}}
    \caption[Reconstructed images using VLA data with \textsc{pdfb}]{Reconstructed images using the \textsc{pdfb} algorithm, shown in $\log_{10}$ scale. Columns denote different dimensionality reduction methods -- from left to right, (i) time- and frequency- averaging, (ii) $\RGt$, and (iii) $\RFPhit$. Rows denote final data sizes achieved after dimensionality reduction -- from top to bottom, (i) $30N\approx2\,000\,000$ visibilities ($\equiv$ full data, no reduction), (ii) $4N\approx 260\,000$ visibilities, (iii) $N\approx 65\,000$ visibilities, (iv) $0.2N\approx 13\,000$ visibilities, (v) $0.05N\approx 3\,200$ visibilities, and (vi) $0.02N\approx 1\,300$ visibilities. Blank spaces in a column represent data sizes that could not be reached with the dimensionality reduction method corresponding to that column.}
    \label{fig:comparemodels}
\end{figure*}

Fig.~\ref{fig:cleanresults} shows image reconstruction using \textsc{ms-clean}\footnote{We used the implementation available in the `WSClean' program~\citep{offringa_wsclean_2014}} on data of varying sizes, obtained by time- and frequency-averaging the initial data set.
The left column shows restored images, where the model, smoothed with the clean beam, is added to the residual image.
We do not show the model images as they are not physically realistic.
\textsc{ms-clean} was run with Briggs weighting (robust weighting parameter set to $-1$) and a major loop gain of 0.8, and took 6 major iterations on the full data set to produce a model image of size $2048\times 2048$ from which a $256\times256$ image was cropped.
Reducing the full data set to lower sizes corresponding to $4N$ and $N$ through a simple time-averaging, over 10 and 20 snapshots respectively, led to increasing artefacts in the reconstructed image.
We can see regular structures in the residuals corresponding to smearing effects in the reconstructed image.
A much lower data size of $0.2N$ was reached by first time-averaging the full data over 30 snapshots and subsequently frequency-averaging over 10 channels.
Running \textsc{ms-clean} on this reduced data led to poor image quality, and may be attributed to the unrealistic averaging needed to reach low data size.
In the first column of Fig.~\ref{fig:comparemodels}, we note that reconstructing images using the \textsc{pdfb} algorithm on the same reduced data sets produced images of better visual quality.
This agrees with results previously reported on data without dimensionality reduction~\citep{onose_scalable_2016,onose_accelerated_2017,vijay_kartik_fourier_2017}.
We can nevertheless observe the adverse effects of drastic averaging methods on \textsc{pdfb}, in the form of artefacts for images reconstructed from very low data sizes, like $0.2N$.

We can thus see that averaging has several limitations as a dimensionality reduction method.
The final data sizes that can be achieved using averaging are limited by the initial number of snapshots and the number of channels in the data set.
The time-averaged data offers a $uv$ coverage that is more incomplete than the original data set and, in addition, the measurement model is inaccurately approximated due to the omission of the averaging operation $\R_{\mathrm{avg}}$.
Consequently, reconstructed images from both \textsc{ms-clean} and \textsc{pdfb} contain related artefacts.
The low data size of $0.2N\approx 13\,000$ visibilities is reached by averaging over arbitrarily large bins.
Reducing data size in this manner is not meaningful, however, since the corresponding loss of information cannot be compensated for by a simple averaging procedure.
Averaging is clearly limited by the need to critically sample the $uv$ plane, and ignoring these hard limits has severe ill-effects on the reconstruction.
Averaging produces reasonable images only if the final data size is much higher than that shown here.
We were able to reconstruct an image with negligible artefacts with a reduced data size of $7N\approx455\,000$ visibilities -- obtained by time-averaging the collated data set over 5 snapshots -- which is much higher than the most conservative reduction performed by $\RGt~(4N)$ or $\RFPhit~(N)$.
Image reconstruction from averaged data may also perform better if the correct measurement model were taken into account, by including the averaging operation $\R_{\mathrm{avg}}$ as mentioned in section~\ref{subsec:averaging}.
The current work, however, mimics the state-of-the-art averaging method which ignores $\R_{\mathrm{avg}}$ at the expense of inaccurate image recovery.

The second and third columns in Figs.~\ref{fig:comparemodels} and \ref{fig:compareresiduals} show a visual comparison of the reconstructed and residual images, respectively, from data reduced using the dimensionality reduction methods $\RGt$ and $\RFPhit$, and then imaged using the \textsc{pdfb} algorithm.
The largest data size that can be achieved after applying $\RGt$ is $4N$, and that for $\RFPhit$ is $N$, which is why target data sizes above these values are shown as blank spaces in the corresponding columns of Figs.~\ref{fig:comparemodels} and \ref{fig:compareresiduals}.
On the other hand, both $\RGt$ and $\RFPhit$ allow us -- by construction -- to attain arbitrarily low data dimensionality by choosing to discard dimensions based either on the significance of the contribution of interpolation kernels to the discrete Fourier grid points (in the case of $\RGt$) or on the significance of the singular values of the original measurement operator (in the case of $\RFPhit$).
We see in Fig.~\ref{fig:comparemodels} that both these dimensionality reduction methods outperform averaging for the same target data sizes.
We note that for final data sizes of approximately the same order as the image size, i.e., $4N, N, 0.2N$, $\RGt$ performs as well as $\RFPhit$.
The robustness of $\RFPhit$, however is apparent when data size is aggressively reduced to as low as $0.05N$ and $0.02N$.
At these extremely low sizes, we can see that data reduced using $\RFPhit$ continue to retain much of the original features of the image (as can be seen in the side lobes in particular) whereas $\RGt$ appears to recover only the overall structure, producing an overly smooth appearance lacking detail.
We note here that the final data size of $0.02N$ is achieved by reducing from an initial data size of $30N$, which represents a dimensionality reduction factor of $\approx1500$, i.e., three orders of magnitude.

Residuals shown in Fig.~\ref{fig:cleanresults} were computed with Briggs weighting using \textsc{ms-clean}.
Residuals shown in Fig.~\ref{fig:compareresiduals} were computed using the original measurement operator for \textsc{pdfb}.
To enable visual comparison of \textsc{clean} and \textsc{pdfb} residual images across columns of Figs.~\ref{fig:cleanresults} and~\ref{fig:compareresiduals}, residual images obtained using \textsc{pdfb} have been scaled by the peak of the point spread function (PSF)\footnote{$\mathrm{PSF_{max}}$, the peak of the instrument response to a point source image at the phase centre with a value $1$ at the central pixel and zero otherwise, i.e., $\mathrm{PSF_{max}} = \max_{\mathrm{i}} (\PPhi^{\dagger}\PPhi\boldsymbol{\delta})_{\mathrm{i}}$, where $\boldsymbol{\delta}$ is the point source image}.
Unsurprisingly, we see an increase in residual structures as we decrease the size of the data used for image recovery (top to bottom).
We note that with $\RFPhit$ (last column of Fig.~\ref{fig:compareresiduals}), we were able to maintain the absence of regular structures down to very low sizes.

An interesting observation is the similarity in the residual images for $\RGt$ for the data sizes $4N$ and $N$.
The corresponding reconstructed images are also very similar to each other.
This may be due to the fact that the initial $uv$ coverage was concentrated in the lower frequencies of the (oversampled) Fourier plane, leaving much of the $uv$ plane empty.
Consequently, the number of \emph{effective} discrete grid points containing contributions from interpolation kernels was much lower than $4N$ ($\approx 0.6N$ in this particular case).
Reducing data size from $4N$ to $N$, therefore, has no effect on the amount of information contained in the `reduced' data since the discarded dimensions would correspond to discrete grid points with zero contributions anyway.
Thus, the reconstructed images look very similar, and a dip in reconstruction quality is only seen when the data size is reduced below the number of effective discrete grid points.

Running \textsc{pdfb} on the full set of visibilities took $\approx$ 2~seconds per iteration. Applying $\RGt$ or $\RFPhit$ reduced the running time of \textsc{pdfb} to $\approx$ 0.2~seconds per iteration.
This may be attributed to the sparse nature of the constituent operators in $\RGt$ and $\RFPhit$.
Additionally, the reduction in data size potentially entails lower memory usage, but this was not directly quantified in our tests.
We see a clear computational advantage of performing dimensionality reduction on the initial data set before invoking the imaging algorithm.
We also note that the quality of images recovered from reduced data produced with $\RFPhit$ and $\RGt$ is comparable to that obtained with the complete set of initial visibilities.

\begin{figure*}
    \centering
    \includegraphics[trim={150px 110px 100px 60px}, clip, width=0.3\linewidth]{{{figs/i256.dl25.k100.n10000.eps6469.gamma3.hyperspec.q6-5-51.allvisibs.resA}}}
    \vspace{0.03in}
    \phantom{\includegraphics[trim={150px 110px 100px 60px}, clip, width=0.3\linewidth]{{{figs/i256.dl25.k100.n10000.eps6469.gamma3.hyperspec.q6-5-51.allvisibs.resA}}}}
    \phantom{\includegraphics[trim={150px 110px 100px 60px}, clip, width=0.3\linewidth]{{{figs/i256.dl25.k100.n10000.eps6469.gamma3.hyperspec.q6-5-51.allvisibs.resA}}}}
    \vspace{0.03in}
    \includegraphics[trim={150px 110px 100px 60px}, clip, width=0.3\linewidth]{{{figs/i256.dl25.k400.n05000.eps6170.gamma5.hyperspec.q6-5-51.averaging.resA}}}
    \includegraphics[trim={150px 110px 100px 60px}, clip, width=0.3\linewidth]{{{figs/i256.dl25.k400.n05000.eps6400.gamma5.hyperspec.q6-5-51.rgrid.resA}}}
	\phantom{\includegraphics[trim={150px 110px 100px 60px}, clip, width=0.3\linewidth]{{{figs/i256.dl25.k400.n05000.eps6400.gamma5.hyperspec.q6-5-51.rgrid.resA}}}}
    \vspace{0.03in}
    \includegraphics[trim={150px 110px 100px 60px}, clip, width=0.3\linewidth]{{{figs/i256.dl25.k100.n05000.eps2013.gamma5.hyperspec.q6-5-51.averaging.resA}}}
    \includegraphics[trim={150px 110px 100px 60px}, clip, width=0.3\linewidth]{{{figs/i256.dl25.k100.n05000.eps6400.gamma5.hyperspec.q6-5-51.rgrid.resA}}}
    \includegraphics[trim={150px 110px 100px 60px}, clip, width=0.3\linewidth]{{{figs/i256.dl25.k100.n05000.eps40.gamma5.hyperspec.q6-5-51.rsing.resA}}}
    \vspace{0.03in}
    \includegraphics[trim={150px 110px 100px 60px}, clip, width=0.3\linewidth]{{{figs/i256.dl25.k020.n05000.eps11038.gamma5.hyperspec.q6-5-51.averaging.resA}}}
    \includegraphics[trim={150px 110px 100px 60px}, clip, width=0.3\linewidth]{{{figs/i256.dl25.k020.n05000.eps6400.gamma5.hyperspec.q6-5-51.rgrid.resA}}}
    \includegraphics[trim={150px 110px 100px 60px}, clip, width=0.3\linewidth]{{{figs/i256.dl25.k020.n05000.eps20.gamma5.hyperspec.q6-5-51.rsing.resA}}}
    \vspace{0.03in}
	\phantom{\includegraphics[trim={150px 110px 100px 60px}, clip, width=0.3\linewidth]{{{figs/i256.dl25.k005.n05000.eps15.gamma5.hyperspec.q6-5-51.rsing.resA}}}}
	\includegraphics[trim={150px 110px 100px 60px}, clip, width=0.3\linewidth]{{{figs/i256.dl25.k005.n05000.eps6400.gamma5.hyperspec.q6-5-51.rgrid.resA}}}
	\includegraphics[trim={150px 110px 100px 60px}, clip, width=0.3\linewidth]{{{figs/i256.dl25.k005.n05000.eps15.gamma5.hyperspec.q6-5-51.rsing.resA}}}
    \vspace{0.03in}
	\phantom{\includegraphics[trim={150px 110px 100px 60px}, clip, width=0.3\linewidth]{{{figs/i256.dl25.k002.n05000.eps15.gamma5.hyperspec.q6-5-51.rsing.resA}}}}
	\includegraphics[trim={150px 110px 100px 60px}, clip, width=0.3\linewidth]{{{figs/i256.dl25.k002.n05000.eps6400.gamma5.hyperspec.q6-5-51.rgrid.resA}}}
    \includegraphics[trim={150px 110px 100px 60px}, clip, width=0.3\linewidth]{{{figs/i256.dl25.k002.n05000.eps15.gamma5.hyperspec.q6-5-51.rsing.resA}}}
    \vspace{0.03in}
    \includegraphics[trim={0px 0px 0px 0px}, clip, width=0.3\linewidth]{{{figs/colourbar_resA}}}
    \includegraphics[trim={0px 0px 0px 0px}, clip, width=0.3\linewidth]{{{figs/colourbar_resA}}}
    \includegraphics[trim={0px 0px 0px 0px}, clip, width=0.3\linewidth]{{{figs/colourbar_resA}}}
    \caption[Residual images using VLA data with \textsc{pdfb}]{Naturally weighted residual images for the \textsc{pdfb} algorithm, shown in linear scale. Columns denote different dimensionality reduction methods -- from left to right, (i) time- and frequency- averaging, (ii) $\RGt$, and (iii) $\RFPhit$. Rows denote final data sizes achieved after dimensionality reduction -- from top to bottom, (i) $30N\approx2\,000\,000$ visibilities ($\equiv$ full data, no reduction), (ii) $4N\approx 260\,000$ visibilities, (iii) $N\approx 65\,000$ visibilities, (iv) $0.2N\approx 13\,000$ visibilities, (v) $0.05N\approx 3\,200$ visibilities, and (vi) $0.02N\approx 1\,300$ visibilities. Blank spaces in a column represent data sizes that could not be reached with the dimensionality reduction method corresponding to that column.}
    \label{fig:compareresiduals}
\end{figure*}

\section{Conclusions}
\label{sec:conclusions}
In this article, we have shown the effectiveness of our proposed dimensionality reduction method, $\RFPhit$, to handle the large volumes of data expected to be acquired in next-generation radio interferometers like the SKA.
It is based on retaining the original information content of the data, and leverages the singular value decomposition of the original measurement operator to achieve this.
An alternative reduction method, $\RGt$, is closely related to the familiar method of `gridding' continuous visibilities to the discrete Fourier grid, and works well when reducing to data sizes close to the image size.
We have shown through Cygnus A image reconstruction using VLA data that both $\RFPhit$ and $\RGt$ outperform the current standard method of reducing data dimension through simple time- and frequency- averaging.
$\RFPhit$ is particularly robust down to extremely low embedded sizes, and is a good candidate for reducing very high-dimensional data.
In our case of reconstructing $256\times256$ size images from well-calibrated VLA data,  a final data size of up to $2$~per~cent of the image size was reached with reasonably low loss in image reconstruction quality.
Given our starting data size of 30 times image size ($30N$), a final data size of 2 per cent of image size ($0.02N$) represents a dimensionality reduction factor of $\approx1500$, i.e., more than three orders of magnitude.
One can expect significantly higher dimensionality reduction ratios for SKA data when the initial data sizes could be many orders of magnitude larger than image size, while the final data size using $\RFPhit$ would always be lower than image size, potentially reaching much lower, depending on $uv$ coverage and other data acquisition characteristics.
In addition to not having the same limitation as averaging methods to reach very low data sizes, $\RFPhit$ and $\RGt$ also produce images with fewer reconstruction artefacts for a comparable data size.
Owing to the modular nature of the constituent operators of $\RFPhit$ and $\RGt$, we propose a mechanism that enables dimensionality reduction to be applied on-the-fly on data as they are being acquired.
This ensures that data size is reduced from the very beginning, thus precluding any issues related to storing or processing large amounts of data in real-time.
This could be a possible addition in the data pipelines for the SKA, which currently estimates handling the massive amounts of data flow to be a serious challenge.
Further work with $\RFPhit$ is foreseen towards addressing calibration issues, and the suitability of dimensionality reduction in the presence of large $w$-terms.
\textsc{matlab} code for the proposed dimensionality reduction method (along with gridding and averaging methods, for comparison) is available on GitHub.

\section*{Acknowledgements}

We would like to thank R.\,Perley from NRAO for providing calibrated VLA data, and for related discussions.
We would also like to thank O.\,M.\,Smirnov for discussions on averaging as a data reduction method.
The work presented in this article was supported by the Swiss National Science Foundation (SNSF) under grant 200020-146594.
AD acknowledges support from the UK Engineering and Physical Sciences Research Council (EPSRC, grants EP/M008843/1).





\bibliographystyle{mnras}
\bibliography{biblio}








\bsp	
\label{lastpage}
\end{document}